\documentclass[11pt]{article}

\usepackage[dvips]{graphicx}

\newcommand{\gev}{\mbox{~GeV}}

\newcommand{\half}{{\textstyle\frac{1}{2}}}

\newcommand{\tvec}[1]{\mbox{\boldmath{$#1$}}}
\newcommand{\svec}[1]{\mbox{\boldmath{$\scriptstyle #1$}}}

\newcommand{\sms}{\mskip 1mu}
\newcommand{\jpsi}{J\mskip -2mu/\mskip -0.5mu\Psi\mskip 0.5mu}

\begin{document}

\begin{flushright}
DESY 05-256 \\
hep-ph/0512201 \\
\end{flushright}

\renewcommand{\thefootnote}{\fnsymbol{footnote}}

\begin{center}
\vspace{1.5\baselineskip}
{\LARGE \bf Generalized parton distributions: recent
  results\,\footnote{Talk presented at the Particles and Nuclei
  International Conference (PANIC 05), Santa Fe, NM, USA, 24--28
  Oct.\ 2005.  To appear in the Proceedings.}}

\vspace{3.0\baselineskip}

M.~Diehl \\[0.5\baselineskip]
\textit{Deutsches Elektronen-Synchroton DESY, 22603 Hamburg, Germany}
\vspace{2.0\baselineskip}
\end{center}

\begin{abstract}
I review progress on selected issues connected with generalized parton
distributions.  Topics range from the description of hard exclusive
reactions to the spatial distribution of quarks in the nucleon and the
contribution of their orbital angular momentum to the nucleon spin.
\end{abstract}

%%%%%%%%%%%%%%%%%%%%%%%%%%%%%%%%%%%%%%%%%%%%

\section{Introduction}

An outstanding task in QCD is to understand hadron structure at the
level of quarks and gluons.  A wealth of information about this
structure is contained in generalized parton distributions (GPDs).
They are related both to the conventional parton densities and to
elastic form factors---quantities which have played a crucial role in
the understanding of strong interactions, and which at first sight are
of very different nature.  Recent reviews on GPDs can be found in
\cite{Goeke:2001tz,Diehl:2003ny}.

\begin{figure}
\begin{center}
  \includegraphics[width=.35\textwidth]{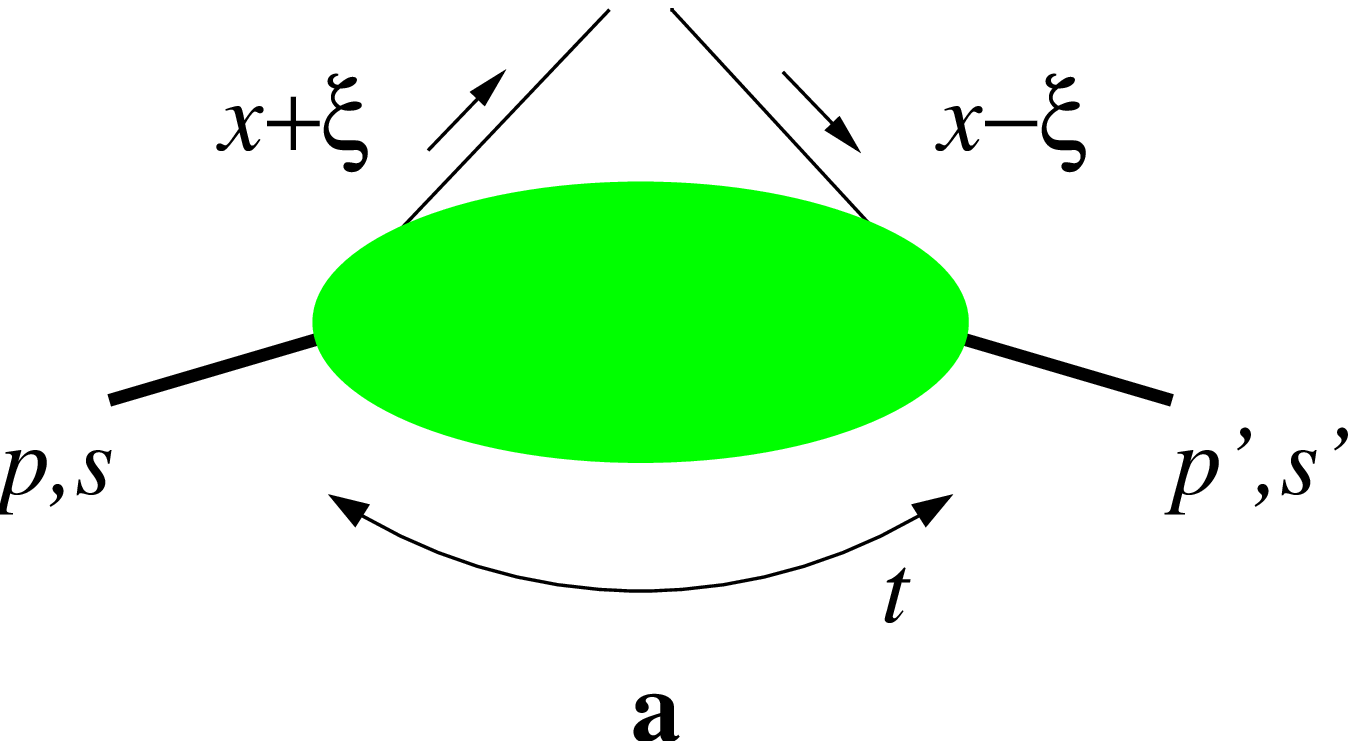}
  \hspace{.1\textwidth}
  \includegraphics[width=.27\textwidth]{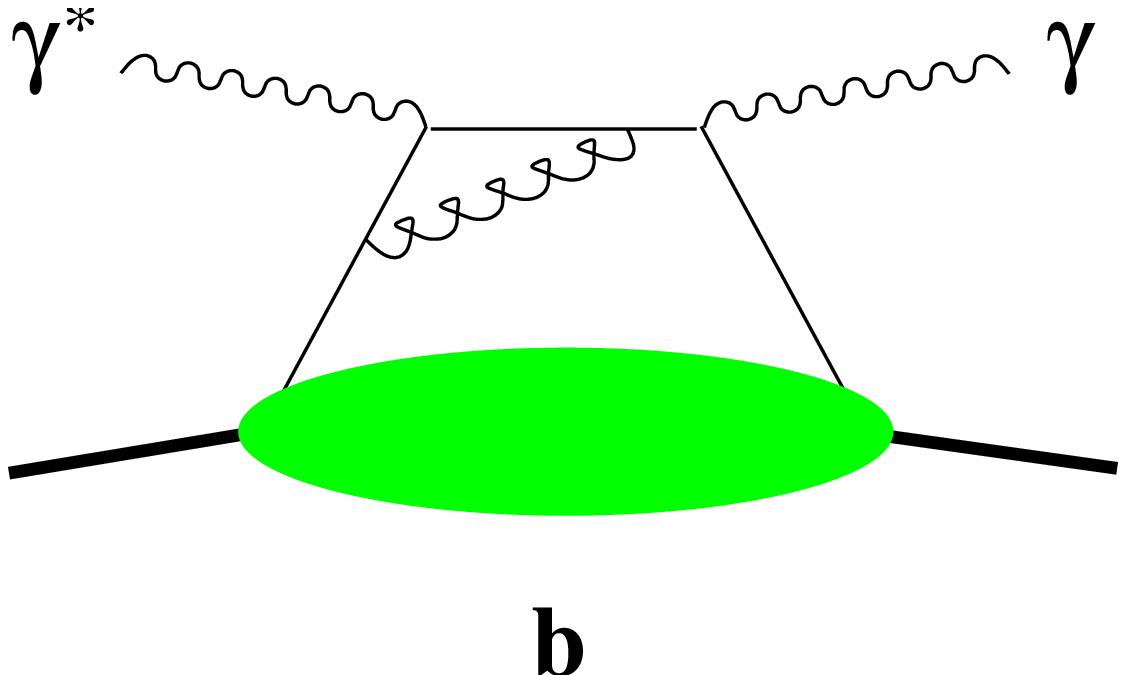}
\end{center}
  \caption{\label{fig:gpd} \textbf{a:} Variables in a GPD.  The
  momentum fractions $x$ and $\xi$ refer to the average hadron
  momentum $\half (p+p')$.  ~~\textbf{b:} A graph for the amplitude of
  deeply virtual Compton scattering (DVCS).}
\end{figure}

GPDs are defined through matrix elements $\langle p' | \mathcal{O} | p
\rangle$ between hadron states $|p'\rangle$ and $|p \rangle$, with
non-local operators $\mathcal{O}$ constructed from quark and gluon
fields.  As sketched in Fig.~\ref{fig:gpd}a, GPDs depend on several
kinematical variables, namely on the longitudinal momentum fractions
$x+\xi$ and $x-\xi$ of the partons and on the invariant momentum
transfer $t= (p-p')^2$.  For unpolarized quarks there are two
distributions $H^q(x,\xi,t)$ and $E^q(x,\xi,t)$.  The former is
diagonal in the proton helicity, whereas the latter describes proton
helicity flip.  For $p=p'$ and equal proton helicities one recovers
the diagonal matrix element parameterized by usual quark and antiquark
densities, so that $H^q(x,0,0)=q(x)$ and $H^q(-x,0,0)=-\bar{q}(x)$ for
$x>0$.  Taking Mellin moments of in $x$, one obtains matrix elements
of \emph{local} operators, which are parameterized by form factors.
In particular, the lowest moments give the well-known Dirac and Pauli
form factors:
\begin{equation}
  \label{ff-sum-rules}
\sum_q e_q \int dx\, H^q(x,\xi,t) = F_1(t) ,
\qquad\qquad
\sum_q e_q \int dx\, E^q(x,\xi,t) = F_2(t) ,
\end{equation}
where $e_q$ denotes the fractional quark charge.  The $\xi$
independence of the integrals is a consequence of Lorentz invariance.
Of great interest is the next highest moment
\begin{equation}
  \label{ji-sum}
\int dx\, x [H^q(x,\xi,t) + E^q(x,\xi,t)] = 2 J^{q}(t) ,
\end{equation}
because $J^{q}(0)$ gives the total angular momentum carried by
quarks and antiquarks of a given flavor, including both their helicity
and their \emph{orbital} angular momentum.  Belonging to local
operators, the Mellin moments of GPDs are well suited for evaluation
in lattice QCD (see below).

\begin{figure}[b]
\begin{center}
  \includegraphics[width=.65\textwidth]{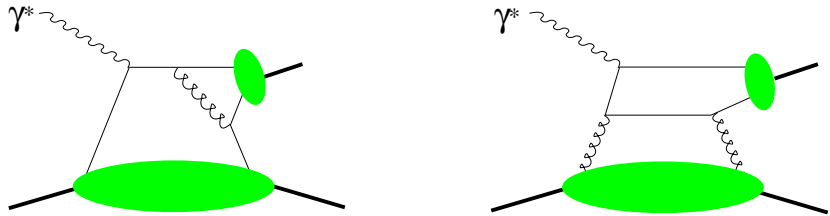}
\end{center}
  \caption{\label{fig:mesons} Example graphs for exclusive meson
  production.  The large blobs represent GPDs and the small ones the
  distribution amplitudes of the produced meson.}
\end{figure}

Factorization theorems state that GPDs appear in the scattering
amplitudes of suitable hard exclusive processes.
Figure~\ref{fig:gpd}b shows an example graph for deeply virtual
Compton scattering, $\gamma^* p\to \gamma p$, where a hard scale is
provided by the initial photon virtuality.  A large number of reaction
channels can be accessed in hard exclusive meson production, such as
$\gamma^* p\to \rho\sms p$ or $\gamma p\to \jpsi\sms p$, where in
addition to quark or gluon GPDs the quark-antiquark distribution
amplitudes of the meson appear as non-perturbative input (see
Fig.~\ref{fig:mesons}).  The longitudinal momentum transfer $\xi$ is
fixed by the process kinematics (for DVCS and light meson production
one simply has $\xi=x_B/(2-x_B)$ in terms of the usual Bjorken
variable).  In contrast, $x$ is a loop variable, and $\xi$ only gives
the \emph{typical} size of $x$ in the convolution of the GPD with the
hard-scattering amplitude.  The hard-scattering subprocesses for
Compton scattering, electroproduction of light mesons, and
photoproduction of heavy quarkonium are fully calculated at
next-to-leading order (NLO) in $\alpha_s$, and partial results for
DVCS at NNLO have just appeared \cite{Muller:2005nz}.  Whereas
$\alpha_s$ corrections are generally found to be moderate for Compton
scattering, they can be substantial for meson production
\cite{Belitsky:2001nq}, and more detailed studies will be needed to
gain quantitative theoretical control over these channels.

In addition to the variables already discussed, GPDs depend on the
scale at which the partons are resolved, given by the hard scale in
the physical process.  The evolution equations for GPDs interpolate
between those for ordinary parton densities and those for meson
distribution amplitudes.  The evolution kernels are known to NLO in
$\alpha_s$, but it is only recently and only to LO accuracy that one
has an explicit solution of the evolution equations
\cite{Kirch:2005tt,Muller:2005ed}.  Generalizing the well-known Mellin
transform technique employed for the ordinary parton densities, this
solution should be useful both for efficient numerical implementation
and for analytic considerations, for instance in the limit of small
$x$.

A typical strategy for modeling GPDs is to take an ordinary parton
density as input and to generate a $\xi$ dependence consistent with
Lorentz invariance relations, which state that Mellin moments in $x$
of GPDs must be polynomials in $\xi$ of a specified degree.  Different
ans\"atze are used to this end, the most common ones being based on
so-called double distributions \cite{Musatov:1999xp} or on the
evolution of GPDs at small $x$ and $\xi$ \cite{Shuvaev:1999ce}.
Different strategies based on moments have recently been proposed in
\cite{Muller:2005ed} and in \cite{Guzey:2005ec}.  With the Compton
amplitude calculated at LO in $\alpha_s$, the model in
\cite{Guzey:2005ec} provides a good description of the data
\cite{Chekanov:2003ya,Aktas:2005ty} for DVCS at small $x_B$.  The $t$
dependence of GPDs will be discussed below.

%%%%%%%%%%%%%%%%%%%%%%%%%%%%%%%%%%%%%%%%%%%%%%%%
\section{Vector meson production}

An important feature of exclusive vector meson production is that
quark and gluon distributions appear at the same order in $\alpha_s$,
whereas in DVCS (just as in inclusive deep inelastic scattering) the
gluon contribution is suppressed by $\alpha_s$ relative to quarks.
Schematically, one has
\begin{equation}
  \label{vector-schem}
\mathcal{A}_{\rho^0} \propto \textstyle \frac{1}{\sqrt{2}} \Big[
  \frac{2}{3} (u+\bar{u}) + \frac{1}{3} (d+\bar{d}) +
  \frac{3}{4} g \Big] ,
\qquad
\mathcal{A}_{\phi} \propto \textstyle \frac{1}{3} (s+\bar{s}) +
  \frac{1}{4} g ,
\end{equation}
where $u+\bar{u}$, $d+\bar{d}$, $s+\bar{s}$, and $g$ represent the
convolutions of the relevant GPDs with the hard-scattering kernels
(which are identical for quarks and gluons at LO).  Taking the
ordinary parton densities as a rough guide for the relative size of
terms in (\ref{vector-schem}), one expects that, even in kinematics of
the fixed-target experiments at HERMES and Jefferson Lab, gluons
provide a substantial contribution to the $\rho^0$ production
amplitude.  This expectation is supported by preliminary HERMES data
for the ratio of $\phi$ and $\rho^0$ production cross sections
\cite{Diehl:2004wj}.  The same conclusion has been reached in an
explicit calculation with GPD models based on double distributions
\cite{Diehl:2005gn}.  In the same analysis, substantial changes in the
calculated cross sections at $Q^2 \sim 4 \gev^2$ and $x_B \sim 0.1$
were found when taking different parameterizations of the gluon
density as input to the GPD model.  This reflects on one hand the
present uncertainty in our knowledge even of the ordinary gluon
distribution, and on the other hand the sensitivity of meson
production to the generalized gluon distribution.  Very similar
findings were made in the study \cite{Tom-Dis:2005} of exclusive
$\jpsi$ production at collider energies: the high-precision data
\cite{Chekanov:2004mw} clearly disfavor a number of parameterizations
of the gluon density in conjunction with the Shuvaev model
\cite{Shuvaev:1999ce} for the $\xi$ dependence of the gluon GPD.

It has long been known that in exclusive meson production the
leading-twist approximation, which gives the leading term in a $1/Q$
expansion, overshoots the data by factors of several for photon
virtualities $Q^2$ below $5 \gev^2$.  Closer analysis reveals that
important corrections to this approximation can be ascribed to the
effect on the hard-scattering amplitude of the intrinsic transverse
momentum of quarks in the produced meson.  A recent analysis
\cite{Goloskokov:2005sd} modeling this effect finds good agreement
with $\rho^0$ and $\phi$ production data at high energies from H1 and
ZEUS.

%%%%%%%%%%%%%%%%%%%%%%%%%%%%%%%%%%%%%%%%%%%%%%%%
\section{Impact parameter distributions}

An important feature of GPDs is that they contain information about
the spatial distribution of quarks and gluons in the nucleon.  This
becomes explicit in the impact parameter representation
\cite{Burkardt:2002hr} (for different approaches see
\cite{Ralston:2001xs} and \cite{Belitsky:2003nz}).  To introduce this
representation, let us form wave packets
\begin{equation}
  \label{impact-state}
|p^+, \tvec{b}\rangle = \int\frac{d^2 p}{(2\pi)^2}\,
  e^{-i\svec{b} \svec{p}}\, |p^+, \tvec{p} \rangle
\end{equation}
from momentum eigenstates $|p^+, \tvec{p} \rangle$, where the
plus-momentum $p^+ = (p^0+p^3) /\sqrt{2}$ simply becomes the
longitudinal momentum (up to a factor $\smash{\sqrt{2}}$) in a frame
where the proton moves fast.  The conjugate variable to the transverse
momentum $\tvec{p}=(p^1,p^2)$ is called impact parameter $\tvec{b}$
and gives the position of the wave packet in the transverse plane.
Indeed, $|p^+, \tvec{b}\rangle$ is an eigenstate of a suitably defined
transverse position operator, i.e., a relativistic particle can be
localized \emph{exactly} in two dimensions (without the ambiguities at
the order of a Compton wavelength occurring when one attempts to
localize a particle in all three dimensions).  A more detailed
analysis shows that $\tvec{b}$ is the ``center of momentum'' of the
partons in the proton, given by a weighted average $\tvec{b} = \sum_i
p_i^+ \tvec{b}_i^{\phantom{+}} /\sum_i p_i^+$ in terms of their
plus-momenta and transverse positions.  The center of momentum is
related by Noether's theorem to a class of Lorentz transformations
called ``transverse boosts'', in analogy to the relation between the
center of mass and Galilean transformations in nonrelativistic
mechanics.

Forming matrix elements from impact parameter states
(\ref{impact-state}) and the operators defining generalized parton
distributions in momentum space, one obtains Fourier transforms of
these distributions.  For $\xi=0$ one finds that
\begin{equation}
  \label{qxb-def}
q(x,b^2) = \int \frac{d^2 \Delta}{(2\pi)^2}\,
  e^{-i \svec{b} \svec{\Delta}}\,
  H^q(x,0,-\tvec{\Delta}^2)
\end{equation}
is the density of quarks with longitudinal momentum fraction $x$ and
transverse distance $\tvec{b}$ from the center of momentum of the
proton.  Integrating $q(x,b^2)$ over $\tvec{b}$ one recovers the usual
quark distribution.  For $\xi\neq 0$ one no longer has a probability
interpretation because the two momentum fractions in
Fig.~\ref{fig:gpd}a are different, but $\tvec{b}$ still describes the
distribution of the struck parton in the transverse plane.  According
to the discussion in the introduction, the combined $\xi$ and $t$
dependence of hard exclusive processes thus gives information about
the impact parameter distribution of partons with longitudinal
momentum fraction of order $\xi$.  Precise measurements are in
particular available for the $t$ dependence of $\jpsi$ production
\cite{Chekanov:2004mw}, giving access to the spatial distribution of
small-$x$ gluons.  H1 has published a first result for the $t$
dependence of DVCS \cite{Aktas:2005ty}, which at the small $x_B$ of
the measurement is sensitive to a combination of sea quark and gluon
distributions.

Like ordinary parton densities, the impact parameter distributions
$q(x,b^2)$ and $g(x,b^2)$ for quarks and gluons depend on the
resolution scale $\mu$.  Their scale evolution is local in $\tvec{b}$
and described by the usual DGLAP equations.  As a consequence, the
$\tvec{b}$ dependence of the distributions at given $x$ also changes
with $\mu$.  A useful quantity to characterize the $\tvec{b}$
distribution is the average squared impact parameter, defined as
\begin{equation}
  \label{av-b}
\langle b^2 \rangle_x
= \frac{\int d^2 b\; b^2\,
  q(x,b^2)}{\int d^2 b\; q(x,b^2)} 
= 4\, \frac{\partial}{\partial t} \log H^q(x,0,t) \Big|_{t=0}
\end{equation}
for quarks and in analogy for gluons.  It is straightforward to obtain
the evolution equations for $\langle b^2 \rangle_x$ from those for the
impact parameter distributions \cite{Diehl:2004cx}.

\begin{figure}
\begin{center}
  \includegraphics[width=0.38\textwidth,%
  bb=0 -55 289 136]{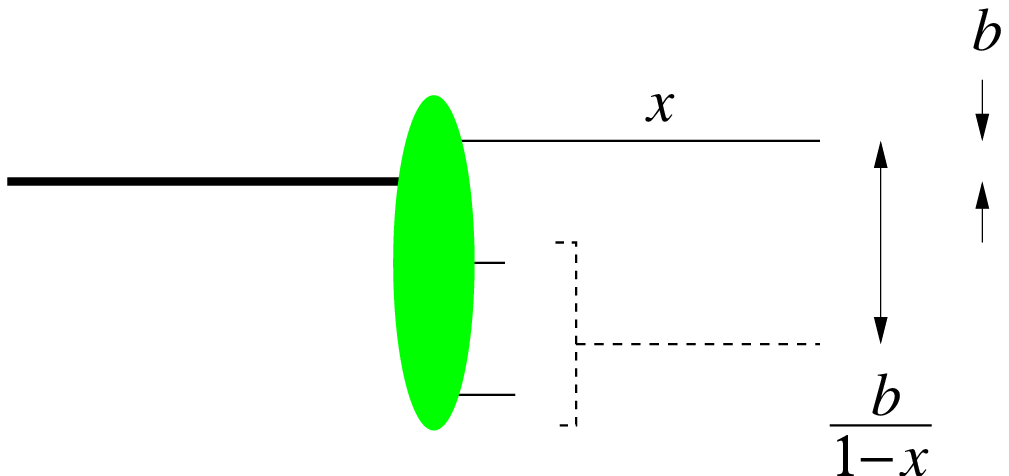}
  \hspace{0.1\textwidth}
  \includegraphics[width=0.4\textwidth]{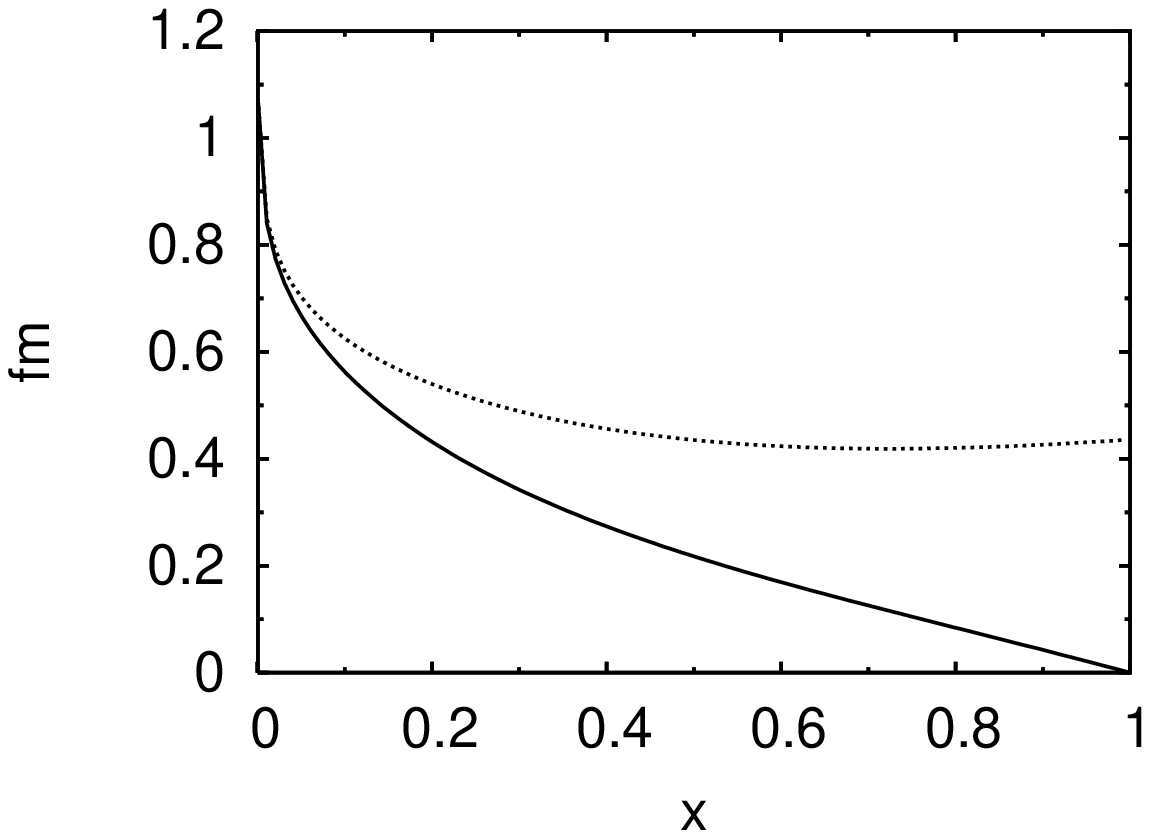}
\end{center}
\caption{\label{fig:impact} {Left:} Three-quark configuration
  with one fast quark in a proton.  The thick line denotes the center
  of momentum of the proton and the dashed line the center of momentum
  of the two spectator quarks.  {Right:} Average impact
  parameter ${\sqrt{\langle b^2 \rangle_x}}$ of the valence $u$
  quark distribution (lower curve) and the associated distance
  $(1-x)^{-1} {\sqrt{\langle b^2 \rangle_x}}$ between the struck
  parton and the center of momentum of the spectators (upper curve),
  as estimated in \protect\cite{Diehl:2004cx}.}
\end{figure}

When a quark takes most of the proton momentum, its impact parameter
will tend to coincide with the center of momentum of the proton as a
whole.  For $x\to 1$ one therefore expects a narrow distribution in
$\tvec{b}$, or equivalently a flat $t$ dependence of GPDs in momentum
space.  An estimate for the overall transverse size of the proton in
that limit is given by the transverse distance $\tvec{b} /(1-x)$
between the struck quark and the center of momentum of the
\emph{spectator} partons, as shown in Fig.~\ref{fig:impact}.  It seems
plausible to assume that this distance remains finite due to
confinement \cite{Burkardt:2004bv}, so that the average squared impact
parameter of partons should vanish like $\langle b^2 \rangle_x \sim
(1-x)^2$ for $x\to 1$.  In the opposite limit of small $x$, the
phenomenology of high-energy hadronic reactions suggests a behavior
like $x^{-(\alpha + \alpha' t)} e^{\,tB}$ of momentum-space GPDs
\cite{Goeke:2001tz,Diehl:2004cx}.  According to (\ref{av-b}) this
translates into a logarithmic growth of the average impact parameter
as $x$ becomes small, $\langle b^2 \rangle_x \sim B + \alpha'
\log(1/x)$.

Indirect information on impact parameter distributions can be obtained
by using the sum rules (\ref{ff-sum-rules}) and the extensive and
precise data on the nucleon form factors
\cite{Diehl:2004cx,Guidal:2004nd}.  For the Dirac form factor $F_1(t)$
this requires an ansatz for the functional dependence of $H^q(x,0,t)$,
which can be restricted to its valence part, since the electromagnetic
current is only sensitive to the difference of quark and antiquark
distributions.  A crucial result of such studies is the rapid decrease
of the average impact parameter with $x$ over the entire $x$ range, as
illustrated for the distribution $u(x,b^2)-\bar{u}(x,b^2)$ in
Fig.~\ref{fig:impact}.  In momentum space, this corresponds to a $t$
dependence that becomes less steep with increasing $x$.  This finding
is confirmed by calculations in lattice QCD, where a clear decrease of
the $t$ slope is seen for moments $\int dx\, x^{n} H^q(x,0,t)$ with
increasing power $n$ \cite{Negele:2004iu}.

\begin{figure}[b]
\begin{center}
  \includegraphics[width=0.34\textwidth]{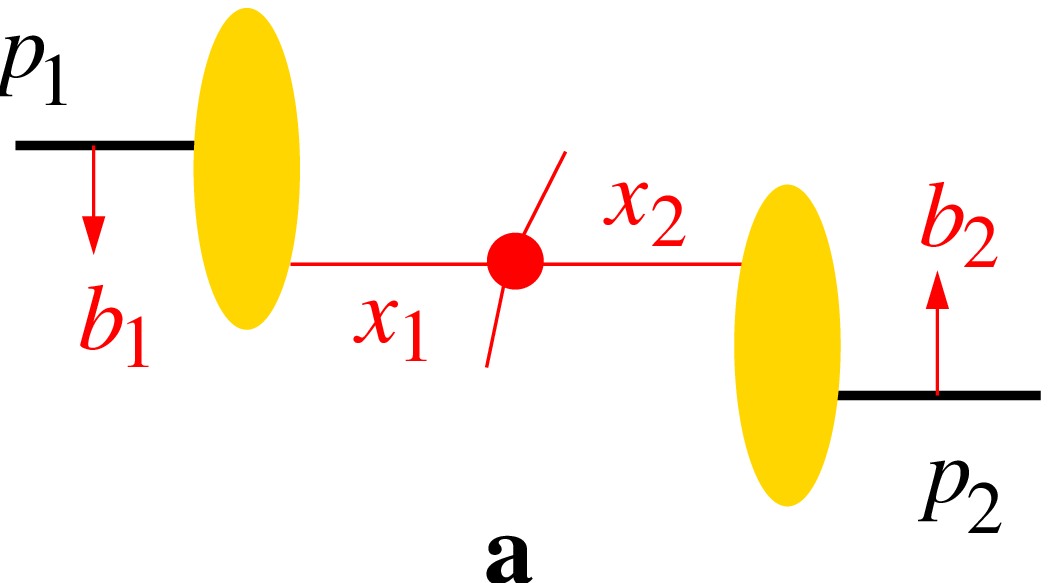}
  \hspace{0.1\textwidth}
  \includegraphics[width=0.34\textwidth]{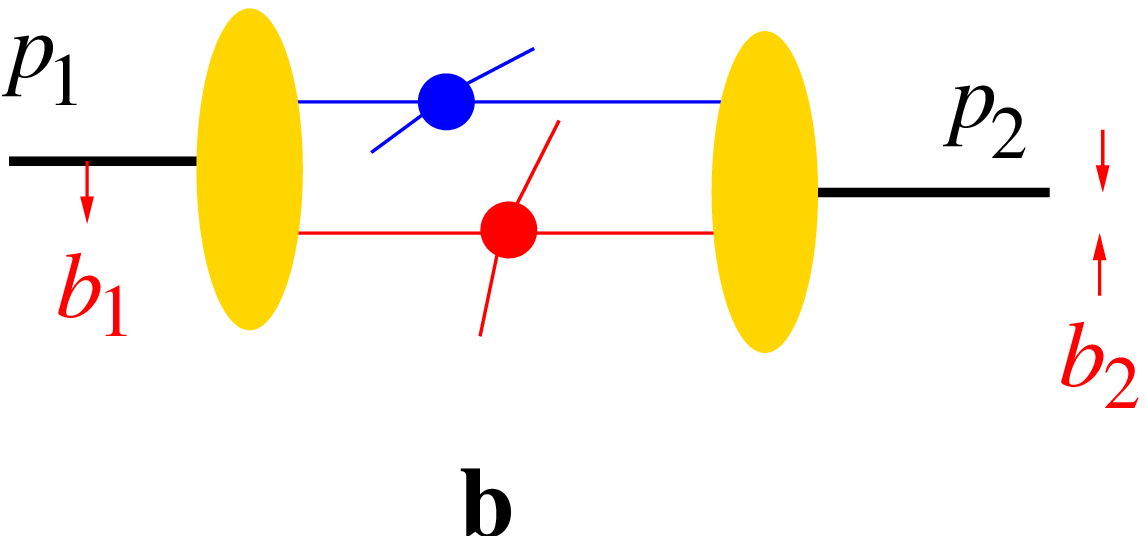}
\end{center}
\caption{\label{fig:multi} \textbf{a:} Graph with a single hard
  interaction in a hadron-hadron collision.  The impact parameters
  $b_1$ and $b_2$ are integrated over independently.  ~~\textbf{b:}
  Graph with a primary and a secondary interaction.}
\end{figure}

A strong correlation between the transverse distribution of partons
and their momentum fraction is not only interesting from the
perspective of hadron structure, but also has practical consequences
for high-energy hadron-hadron collisions \cite{Frankfurt:2003td}.
Consider the production of a high-mass system (e.g.\ a dijet or a
heavy particle).  For the inclusive production cross section, the
distribution of the colliding partons in impact parameter is not
important: only the parton distributions integrated over impact
parameters are relevant according to standard hard-scattering
factorization (see Fig.~\ref{fig:multi}a).  There can however be
additional interactions in the same collision, especially at the high
energies of the Tevatron or the LHC, as shown in
Fig.~\ref{fig:multi}b.  Their effect cancels in sufficiently
\emph{inclusive} observables, but it does affect the event
characteristics and can hence be quite relevant in practice.  In this
case, the impact parameter distribution of partons does matter: the
production of a heavy system requires large momentum fractions for the
colliding partons.  A narrow impact parameter distribution for these
partons forces the collision to be more central, which in turn
increases the probability for multiple parton collisions in the event.

%%%%%%%%%%%%%%%%%%%%%%%%%%%%%%%%%%%%%%%%%%%%%%%%
\section{Spin and orbital angular momentum}

The proton helicity-flip distributions $E^q$ have a density
interpretation at $\xi=0$, similar to the distributions $H^q$
discussed so far.  To see this one changes basis from proton helicity
states $|\!\!\uparrow \rangle$, $|\!\!\downarrow \rangle$ to states
$|X\pm \rangle = (\, |\!\!\uparrow \rangle \pm |\!\!\downarrow \rangle
\,) /\sqrt{2}$ polarized along the positive or negative $x$ axis.  In
impact parameter space one then obtains the density
\begin{equation}
  \label{trans-density}
q_X(x,\tvec{b}) = q(x,b^2) 
  - \frac{b^y}{m} \frac{\partial}{\partial b^2}\,
    e^{\,q}(x,b^2)
\end{equation}
of unpolarized quarks in a proton polarized along the positive $x$
direction, where the Fourier transform
\begin{equation}
e^{\,q}(x,b^2) = \int \frac{d^2 \Delta}{(2\pi)^2}\,
  e^{-i \svec{b} \svec{\Delta}}\,
  E^q(x,0,-\tvec{\Delta}^2)
\end{equation}
is defined in analogy to (\ref{qxb-def}) and $m$ denotes the nucleon
mass.  The impact parameter distribution of quarks in a transversely
polarized proton is thus shifted in the direction perpendicular to the
polarization.  This shift must be substantial at least for some $x$
and $\tvec{b} \sms$: due to the sum rule (\ref{ff-sum-rules}) the
average $\int dx\, \int d^2 b\, e^{\,q}(x,b^2) = \kappa^{\,q}$ is
given by the magnetic moments of proton and neutron as $\kappa^u
\approx 1.67$ and $\kappa^d \approx -2.03$ and hence quite large.  A
connection has been proposed in \cite{Burkardt:2003je} between the
anisotropy (\ref{trans-density}) in the \emph{spatial} distribution of
quarks and the Sivers effect, which is an anisotropic \emph{transverse
momentum} distribution of quarks in a transversely polarized proton.
The Sivers effect has indeed been observed experimentally
\cite{Airapetian:2004tw}.  Similar asymmetries can be discussed for
the distribution of transversely polarized quarks in an unpolarized
proton \cite{Diehl:2005jf}, and first results in lattice QCD indicate
that the corresponding anisotropy of the impact parameter distribution
is appreciable \cite{Diehl:2005ev}.

\begin{figure}
\begin{center}
\includegraphics[width=0.28\textwidth]{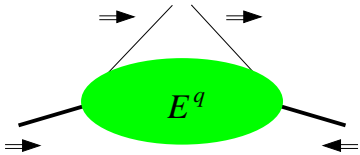}
\end{center}
\caption{\label{fig:gpdE} A transition where the proton helicity is
   flipped but the quark helicity conserved.  The helicity mismatch is
   compensated by one unit of orbital angular momentum, so that
   angular momentum is conserved.}
\end{figure}

The interpretation of (\ref{trans-density}) as a density implies
positivity conditions for $e^{\,q}(x,b^2)$ \cite{Burkardt:2003ck}.  In
momentum space one has a simple bound
\begin{equation}
  \label{e-bound}
| E^q(x,0,0) | \le q(x)\, m \sqrt{\langle b^2 \rangle_x} \; ,
\end{equation}
with more stringent inequalities involving also quark helicity
dependent distributions on the right-hand side.  According to the
behavior of $\langle b^2 \rangle_x$ discussed in the previous section,
this restricts $E^q$ quite severely at larger values of $x$.  The
distribution $E^q$ involves one unit of orbital angular momentum since
in the associated matrix elements the proton helicity is flipped but
the quark helicity conserved (see Fig.~\ref{fig:gpdE}).  The bound
(\ref{e-bound}) thus limits the amount of orbital angular momentum
that can be carried by quarks \cite{Burkardt:2005km}, and is
especially strong at large $x$.

Further information on the helicity-flip distribution can be obtained
from the sum rule for the Pauli form factor in (\ref{ff-sum-rules}).
For this one needs an ansatz for the valence part of $E^q(x,0,t)$,
which can be made in analogy to the case of $H^q(x,0,t)$ and the Dirac
form factor \cite{Diehl:2004cx,Guidal:2004nd}.  The corresponding fits
are less well constrained, because in contrast to $H^q$ the forward
limit $E^q(x,0,0)$ is not known and needs to be parameterized in
addition to $t$ dependence of $E^q$.  It turns out that the positivity
bound (\ref{e-bound}) and its stronger versions provide valuable
restrictions of the allowed parameter space.  One can use these
restrictions and estimate the total angular momentum $J_v^{q}$ carried
by valence quarks, taking the analog of Ji's sum rule (\ref{ji-sum})
for the difference of quark and antiquark contributions.  The result
obtained in \cite{Diehl:2004cx,Diehl:2005wq} is $2J_v^u = 0.39 \div
0.46$ and $2J_v^d = -0.04 \div 0.04$ at scale $\mu=2 \gev$.
Subtracting the quark helicity parts, one gets rather large
numbers $2L_v^u = - (0.47 \div 0.54)$ and $2L_v^d = 0.30 \div
0.38$ for the orbital angular momentum carried by $u$ and $d$ valence
quarks.  The corresponding estimate for the isovector combination
$2(L_v^u - L_v^d) = - ( 0.77 \div 0.92)$ is large, whereas the
isoscalar combination $2(L_v^u + L_v^d) = - (0.11 \div 0.22)$ is
uncertain and quite small.  Lattice calculations by the QCDSF
Collaboration obtain $2(L^u - L^d) = 0.90\pm 0.12$ and $L^u + L^d$
compatible with zero within errors \cite{GerritDis:2005}.  Given the
very different sources of systematic errors in the lattice evaluation
and the estimate from \cite{Diehl:2004cx,Diehl:2005wq}, the agreement
between the two is encouraging, especially for the isovector
combination, where sea quark contributions should be small.

The angular momentum carried by sea quarks is hardly known.  It cannot
be inferred form electromagnetic form factors, and in lattice QCD it
requires calculation of so-called ``disconnected'' graphs, which are
affected by large statistical errors.  Information on sea quarks can
however be obtained from hard exclusive processes, especially from
DVCS, where the control of theory is highest and a large number of
spin asymmetries can be evaluated in the leading-twist approximation.
HERMES has presented preliminary results for the spin asymmetry in
DVCS on a transversely polarized proton \cite{Ye:2005pf}.  Their
comparison with a calculation \cite{Ellinghaus:2005uc} using $J^q$ as
input parameter in the ansatz for $E^q$ shows that this asymmetry is
indeed sensitive to $J^u$ in the kinematics of the experiment (recall
that in Compton scattering on the proton $u$ and $d$ quark
distributions have relative weight $4:1$ due to the squared quark
charges).  A measurement of DVCS on the neutron at Jefferson Lab Hall
A \cite{JLAB:E03-106} is currently being analyzed and should provide
information about $J^d$.  A number of ongoing measurements at
Jefferson Lab and DESY can provide information on the unpolarized
distributions $H^q$ and $H^g$.  Future experimental prospects for this
field are a proposed run of COMPASS at CERN with a detector for
recoiling protons \cite{DHose:2002ia}, the planned upgrade of
Jefferson Lab to $12 \gev$, and eventually an electron-proton collider
eRHIC/EIC.

%%%%%%%%%%%%%%%%%%%%%%%%%%%%%%%%%%%%%%%%%%%%%%%%
\section{Conclusions}

In this talk I have reviewed progress in selected areas connected with
generalized parton distributions.  There are many interesting aspects
discussed in the recent literature which I could not cover for reasons
of time.  Among these are studies of GPDs in dynamical models, GPDs of
nuclei, production of exotic mesons, GPDs for hadron-photon and
baryon-meson transitions, hard exclusive processes at large $s$ and
$t$, and possibilities to study GPDs in neutrino-nucleon scattering
with future high-intensity neutrino beams.

To summarize, there has been important technical progress in the
description of hard exclusive processes, with full NLO results in
$\alpha_s$ available for most relevant channels, partial NNLO results
for Compton scattering, and a better understanding of the scale
evolution of GPDs.  These advances remain to be fully implemented in
the analysis of data, but existing studies have in particular shown
the high sensitivity of vector meson production to the generalized
gluon distribution, even at the moderate $x$ accessible in
fixed-target experiments.

Through the impact parameter representation, GPDs provide information
on the spatial distribution of partons in a hadron, and a number of
studies have turned this concept into quantitative information.
Phenomenological analysis of elastic form factors, as well as results
from lattice QCD show that the average impact parameter of valence
quarks strongly decreases with their momentum fraction in the proton.
The proton helicity-flip distribution $E^q$ has connections with two
crucial aspects of spin physics: transverse polarization effects and
the orbital angular momentum $L^q$ carried by quarks in the nucleon.
First steps have been taken towards a quantitative understanding of
$L^q$.  The present picture---suggested both by lattice results and
model-dependent analysis of experimental data---is that for individual
quark flavors $L^q$ may be substantial, but that when summed over
flavors the orbital angular momentum carried by \emph{valence} quarks
contributes little to the nucleon spin.  The proton spin puzzle thus
remains a puzzle.

%%%%%%%%%%%%%%%%%%%%%%%%%%%%%%%%%%%%%%%%%%%%%%%%

\section*{Acknowledgments}
I thank the Conference Organizers for the kind invitation to attend
this very fruitful meeting.  This work is supported by the Helmholtz
Association, contract number VH-NG-004.

\end{document}